\documentclass[journal, 10pt, twocolumn]{IEEEtran}
\usepackage{cite}
\usepackage{amsmath,amssymb,amsfonts}
\usepackage{algorithmic}
\usepackage{graphicx}
\usepackage{textcomp}
\usepackage{graphics}
\usepackage{epsfig}
\usepackage{epsf}
\usepackage{float}

\newcommand{\norm}[1]{\lVert#1\rVert}
\begin{document}

\title{New Constellation Design and Bit Mapping for Dual Mode OFDM-IM}
\author{Kee-Hoon Kim
\thanks{Department of Electronic Engineering, Soonchunhyang University, Asan 31538, Korea (e-mail: keehk85@gmail.com).}}

\maketitle
\begin{abstract}
Dual mode orthogonal frequency division multiplexing with index modulation (DM-OFDM-IM) is recently proposed, which modulates all subcarriers to eliminate the limits of spectrum efficiency in OFDM with index modulation (OFDM-IM). In DM-OFDM-IM, the subcarriers within each subblock are
divided into two groups, which are modulated by two distinguishable constellation alphabets drawn from the distinct two constellations.

In this paper, a new constellation design for DM-OFDM-IM is proposed based on the pairwise error probability (PEP) analysis. In previous works, the design of the constellation pair is naive and induces much power. Also, a new bit mapping scheme for DM-OFDM-IM is proposed. The proposed bit mapping scheme can relieve the catastrophic effect of the index demodulation error incurring burst of errors, which is the endemic problem of the index demodulation systems. The simulation results show that the proposed constellation pair and the proposed bit mapping scheme substantially enhance the BER performance of the DM-OFDM-IM than the conventional DM-OFDM-IM.
\end{abstract}


\section{Introduction}
In recent years, multicarrier transmission has become an attractive technique in many wireless standards to meet the increasing demand for high data rate communication systems. One of the most popular multicarrier techniques, orthogonal frequency division multiplexing (OFDM), has developed
into a widely-used scheme for wideband digital communication. The major advantage of OFDM over single-carrier
schemes is its ability to cope with frequency-selective fading channel with only one-tap equalizer.
OFDM has been the most popular multicarrier transmission technique in wireless
communications and has become an integral part of IEEE 802.16 standards \cite{andrews2014will}.
Many attempts to further improve the classical OFDM system have been made.

The concept of index modulation (IM) originates from spatial modulation technique \cite{mesleh2008spatial,di2014spatial,yang2015design} in multiple-input multiple-output (MIMO) systems. It utilizes the indices of antennas to convey extra information during the transmission. IM was introduced into OFDM systems as subcarrier index modulation (SIM) in \cite{abu2009subcarrier}, where additional bits can be transmitted through subcarrier indices. Spectral efficiency and energy efficiency can be improved at the same time since it exploits another dimension for transmission.
By motivated this work, several OFDM with index modulation (OFDM-IM) schemes have been proposed in \cite{tsonev2011enhanced,fan2014orthogonal,bacsar2013orthogonal}. However, in common index modulation systems, a portion of subcarriers are deactivated and unused during the
transmission, which results in the decrease of the overall spectral efficiency.

More recently, dual mode OFDM with index modulation (DM-OFDM-IM) has been introduced in \cite{mao2017dual}. The DM-OFDM-IM system modulates all subcarriers using an independent pair of distinct two constellations and implicitly transmits additional bits through the subcarrier indices pattern. In such fashion, DM-OFDM-IM has the
potential to further accelerate the data transmission rate. Similar to OFDM-IM, a generalized version of DM-OFDM
was proposed in \cite{mao2017generalized}. However, the constellation pair used in \cite{mao2017dual} and \cite{mao2017generalized} are naively designed and induces a huge power increase. Also, the erroneously detected subcarrier pattern induces error propagation and will incur burst of bit errors in modulated symbols, which is inherited from the OFDM-IM systems.

In this paper, there are two contributions as follows.
\begin{itemize}
  \item A new constellation pair for DM-OFDM-IM is proposed. The proposed constellation pair is designed based on the pairwise error probability (PEP) analysis. As a result, the DM-OFDM-IM using the proposed constellation pair shows a better BER performance than the DM-OFDM-IM using the conventional constellation pair in \cite{mao2017dual}.
  \item A new bit mapping scheme for DM-OFDM-IM is proposed. The proposed bit mapping scheme can enhance the robustness of the index modulation in DM-OFDM-IM. In the proposed bit mapping scheme, the incoming bit stream generates the modulated symbols in order from the front regardless of the constellation they belong to. By doing this, the catastrophic effect of the index demodulation error incurring burst of errors can be relieved. To the authors' best knowledge, the proposed bit mapping is the first work to solve that problem, which is the endemic drawback of the index modulation based systems.
\end{itemize}
The benefit of two proposed features is verified through simulations, where DM-OFDM-IM using the proposed constellation pair shows a better BER performance than the DM-OFDM-IM using the conventional constellation pair. Also, the new bit mapping scheme provices a better BER performance than the conventional bit mapping.

The rest of this paper is organized as follows. Section II describes the system model of DM-OFDM-IM and the conventional constellation pair. Section III presents the proposed constellation pair based on PEP analysis. Also, the theoretical comparison between the conventional and proposed constellation pair is given. Section IV proposes the new bit mapping scheme for DM-OFDM-IM. In Section V, the simulation results are given to evaluate the benefit of the proposed two features for DM-OFDM-IM. Finally, our conclusions are drawn in Section VI.

\subsection{Notations}
Vectors are denoted by boldface letters as $\mathbf{X}$ and its $\alpha$-th element is denoted by $X(\alpha)$.
$\mathcal{CN}(0,\sigma^2)$ represents the distribution of circularly symmetric complex Gaussian (CSCG) random variable with zero mean and variance $\sigma^2$. The Euclidean norm of a vector $\mathbf{X}$ is denoted by $\norm{\mathbf{X}}_2$. Also, $\hat{I}_A$ and $\hat{\mathbf{X}}$ mean the estimate of a set $I_A$ and a vector $\mathbf{X}$, respectively.

\section{Dual Mode OFDM-IM and the Conventional Constellation Pair}
\subsection{Dual Mode OFDM-IM \cite{mao2017dual}}\label{sec:DM-OFDM-IM}

\begin{figure}
  \centering
  \includegraphics[width=.95\linewidth]{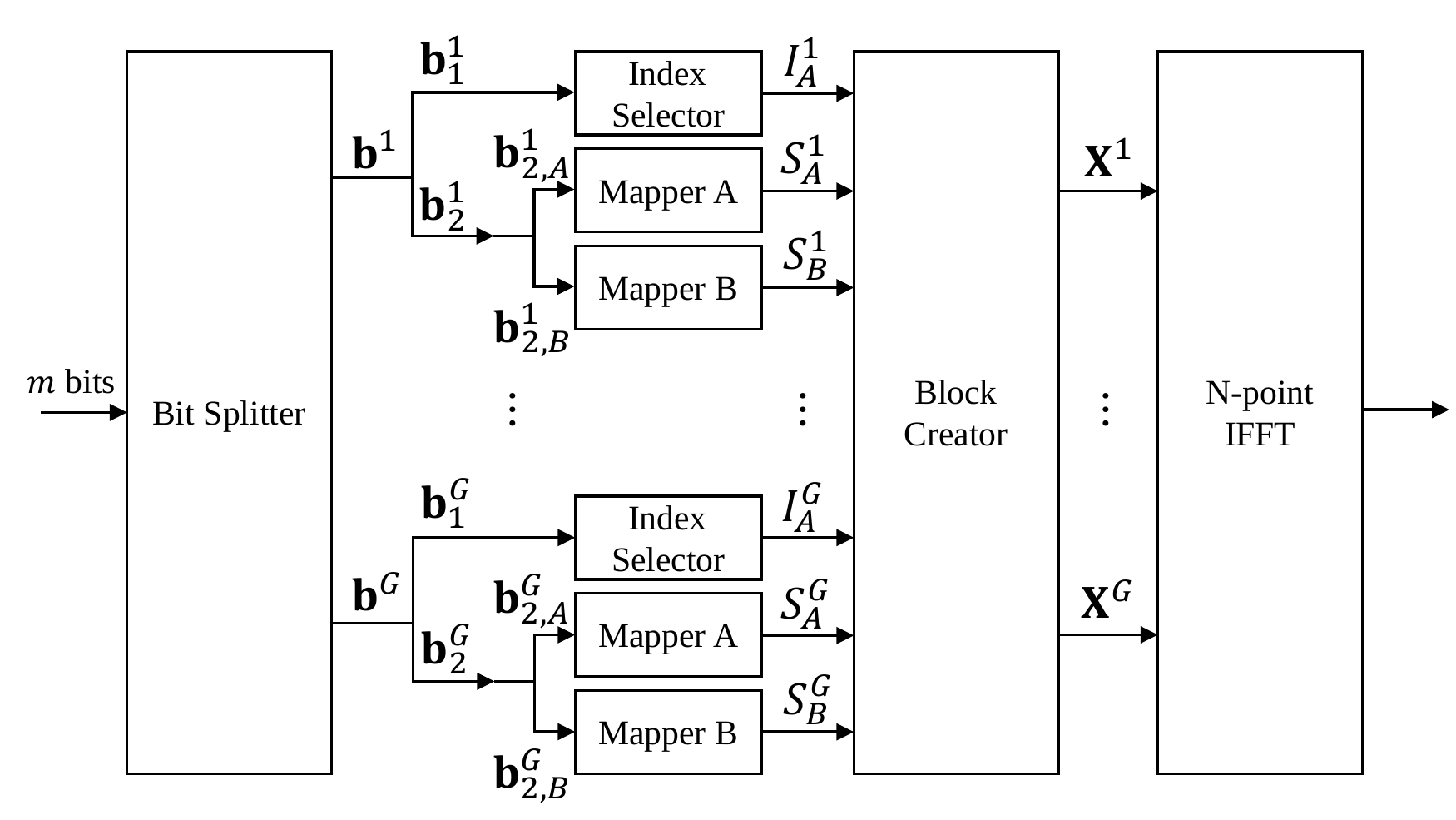}
  \caption{The block diagram of the DM-OFDM-IM transmitter.}\label{fig:bd}
\end{figure}

The DM-OFDM-IM transmitter is illustrated in Fig. \ref{fig:bd}. First, $m$ incoming bits are partitioned by a bit splitter into $G$ groups, $\mathbf{b}^1,\cdots,\mathbf{b}^G$, where each consisting of $p$ bits as
\begin{equation*}
  \mathbf{b}^g = [b^g(1)~b^g(2)~\cdots~b^g(p)],~~~g=1,\cdots,G
\end{equation*}
and $p=m/G$.
Each bit stream $\mathbf{b}^g$ is fed into an index selector and two different constellation mappers for generating a group of length $n=N/G$, where $N$ is the size of fast Fourier transform (FFT). In contrast to the existing OFDM-IM, whereby only part of the subcarriers
are actively modulated, in the DM-OFDM-IM scheme all the subcarriers are modulated in each group, which leads to an enhanced spectral efficiency.
That is, $k$ out of $n$ subcarriers are modulated by $M_A$-ary modulation from the first constellation $\mathcal{M}_A$ and the remaining $n-k$ subcarriers are modulated by $M_B$-ary modulation from the seconde constellation  $\mathcal{M}_B$ in each group. We denote the constellation pair as $(\mathcal{M}_A,\mathcal{M}_B)$ and $M_A$ can be different from $M_B$.

Prior to explaining the DM-OFDM-IM, we represent the bit stream for generating $g$-th group as
\begin{align*}
\mathbf{b}^g =& [\mathbf{b}_1^g~\mathbf{b}_2^g]\\
             =& [\mathbf{b}_1^g~\mathbf{b}_{2,A}^g~\mathbf{b}_{2,B}^g]
\end{align*}
where
\begin{equation*}
\mathbf{b}_2^g = [\mathbf{b}_{2,A}^g~\mathbf{b}_{2,B}^g]
\end{equation*}
and
\begin{align*}
  &\mathbf{b}_1^g = [b^g(1)~\cdots~b^g(p_1)] \\
  &\mathbf{b}_{2,A}^g = [b^g(p_1+1)~\cdots~b^g(p_1+p_{2,A})] \\
  &\mathbf{b}_{2,B}^g = [b^g(p_1+p_{2,A}+1)~\cdots~b^g(p_1+p_{2,A}+p_{2,B})]
\end{align*}
whose sizes are $p_1$, $p_{2,A}$, and $p_{2,B}$, respectively. Clearly, the size of $\mathbf{b}_2^g$ becomes $p_2 = p_{2,A} + p_{2,B}$.

From $\mathbf{b}^g$, the first $p_1$ bit stream $\mathbf{b}_1^g$ is utilized by the index selector. It determines the indices of the subcarriers modulated by $\mathcal{M}_A$ in the $g$-th group. We denote the corresponding index set as
\begin{equation}
  I_A^g = \{i_{A,1}^g,\cdots,i_{A,k}^g\}
\end{equation}
where $i_{A,j}^g \in \{1,2,\cdots,n\}$ and $j = 1,\cdots,k$. Then, the indices of the subcarriers modulated by $\mathcal{M}_B$ in the $g$-th group is uniquely determined by $I_A^g$. We denote the corresponding index set as \begin{equation}
  I_B^g = \{i_{B,1}^g,\cdots,i_{B,n-k}^g\}
\end{equation}
where $i_{B,u}^g \in \{1,2,\cdots,n\}$ and $u = 1,\cdots,n-k$.
The possible pattern of $I_A^g$ is $\binom{n}{k}$ and thus $p_1$ is
\begin{equation*}
  p_1 = \left\lfloor \log_2 \binom{n}{k}\right\rfloor
\end{equation*}

The next $p_{2,A}$ bit stream $\mathbf{b}_{2,A}^g$ is passed to the mapper A and generate the set of $k$ modulated symbols by $\mathcal{M}_A$ in the $g$-th group, which is represented as
\begin{equation}
  S_{A}^g = \{s_{A,1}^g,\cdots,s_{A,k}^g\}
\end{equation}
where $s_{A,j}^g \in \mathcal{M}_A$ and $j = 1,\cdots,k$.
Accordingly, we have
\begin{equation*}
  p_{2,A} = k\log_2 M_A
\end{equation*}

The last $p_{2,B}$ bit stream $\mathbf{b}_{2,B}^g$ is passed to the mapper B and generate the set of $(n-k)$ modulated symbols by $\mathcal{M}_B$ in the $g$-th group, which is represented as
\begin{equation*}
  S_{B}^g = \{s_{B,1}^g,\cdots,s_{B,n-k}^g\}
\end{equation*}
where $s_{B,u}^g\in \mathcal{M}_B$ and $u = 1,\cdots,n-k$.
Accordingly, we have
\begin{equation*}
  p_{2,B} = (n-k) \log_2 M_B
\end{equation*}
Clearly, we have to design the constellation pair $(\mathcal{M}_A,\mathcal{M}_B)$ satisfying $\mathcal{M}_A \cap \mathcal{M}_B = \varnothing$.

As a result, the DM-OFDM-IM block for $g$-th group is formed according to $I_{A}^g$, $S_{A}^g$, and $S_{B}^g$ as
\begin{equation}\label{eq:Xg}
  \mathbf{X}^g = [X^g(1)~X^g(2)~\cdots~X^g(n)]
\end{equation}
where $X^g(i_{A,j}^g) = s_{A,j}^g,j=1,\cdots,k$ and $X^g(i_{B,u}^g) = s_{B,u}^g,u=1,\cdots,n-k$
After the formation all $G$ groups, they are concatenated to generate the DM-OFDM-IM symbol sequence as
\begin{equation}
  \mathbf{\bar{X}} = [\mathbf{X}^1~\mathbf{X}^2~\cdots~\mathbf{X}^G].
\end{equation}
Then, it is transformed into DM-OFDM-IM signal sequence in time domain by inverse FFT (IFFT) and transmitted.

At the receiver, the frequency domain received symbol sequence of $g$-th group is
\begin{equation}\label{eq:YXHW}
  \mathbf{{Y}^g} = \mathbf{{X}^g}\mathbf{{H}^g} + \mathbf{{W}^g}
\end{equation}
where $\mathbf{{Y}^g} = [Y^g(1)~\cdots~Y^g(n)]$ is the received symbol sequence, $\mathbf{{H}^g} = \mathrm{diag}([H^g(0)~H^g(1)~\cdots~H^g(n)])$ is the $n \times n$ matrix whose diagonal elements are channel frequency response (CFR), $\mathbf{W}^g = [W^g(1)~W^g(2)~\cdots~W^g(n)]$ is the additive white Gaussian noise (AWGN) for $g$-th group. The noise level is $W^g(\alpha) \sim \mathcal{CN}(0, N_0),\alpha = 1,\cdots,n$.
Note that the performance within different groups are identical and it is sufficient to investigate a single group to determine the overall system performance \cite{bacsar2013orthogonal}.

The optimal maximum-likelihood detector for the $g$-th group is
\begin{equation}\label{eq:ML}
  \hat{\mathbf{X}}^g = \arg\min_{\mathbf{X}^g}   \norm{\mathbf{Y}^g-\mathbf{X}^g \mathbf{H}^g}_2
\end{equation}
where $\mathbf{X}^g$ is the possible realizations using possible ${I}_{A}^g$, ${S}_{A}^g$, and ${S}_{B}^g$. Clearly, the number of possible realizations of $\mathbf{X}^g$ is $2^{p_1}M_A^k M_B^{n-k}$.
From $\hat{\mathbf{X}}^g$, the estimates $\hat{I}_{A}^g$, $\hat{S}_{A}^g$, and $\hat{S}_{B}^g$ are obtained and, from these, the estimates of bit streams $\hat{\mathbf{b}}_1^g$, $\hat{\mathbf{b}}_{2,A}^g$, and $\hat{\mathbf{b}}_{2,B}^g$ are obtained, respectively.

It can be seen from (\ref{eq:ML}) that the computational complexity of the ML detector in terms of complex multiplications is on the order of $2^{p_1}M_A^k M_B^{n-k}$ per group, denoted as $\mathcal{O}(2^{p_1}M_A^k M_B^{n-k})$. Therefore, the ML detector is impractical to implement for large $p_1$, $n$, $k$ and modulation orders $M_A$ and $M_B$, due to its exponentially
increasing complexity. In \cite{zhang2017dual}, a low complexity ML detector is proposed, which requires complex multiplications on the order of $\mathcal{O}(n (M_A + M_B))$, which can be implemented practically. Therefore, we use this low complexity ML detector in simulations.

\subsection{The Conventional Signal Constellation Pair}

\begin{figure}
  \centering
  \includegraphics[width=.7\linewidth]{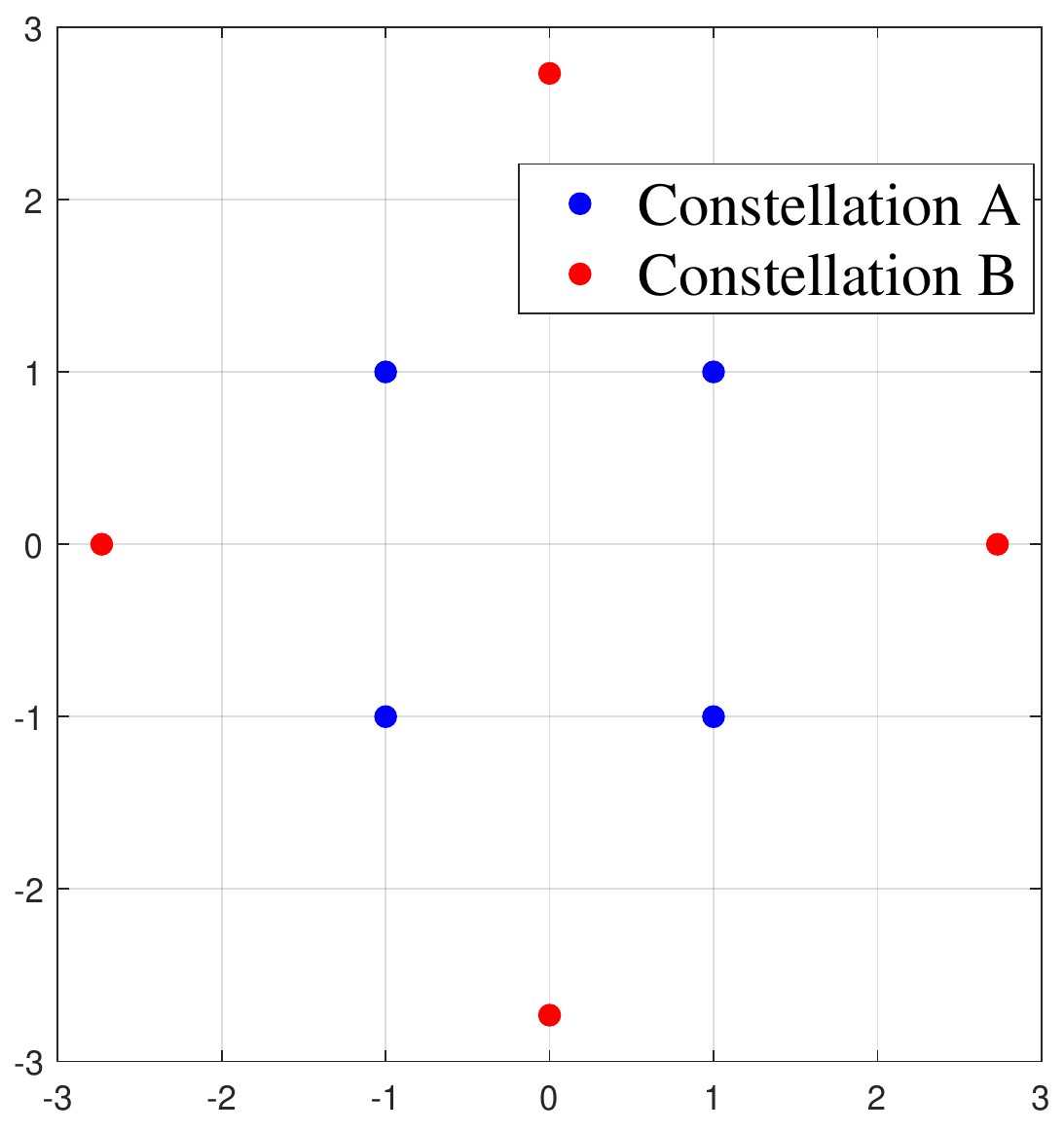}
  \caption{The conventional constellation pair $\mathcal{M}_A^{\mathrm{Conv,QPSK}}$ (Constellation A) and $\mathcal{M}_B^{\mathrm{Conv,QPSK}}$ (Constellation A) when $M_A = M_B = 4$.}\label{fig:CQPSK}
\end{figure}
\begin{figure}
  \centering
  \includegraphics[width=.7\linewidth]{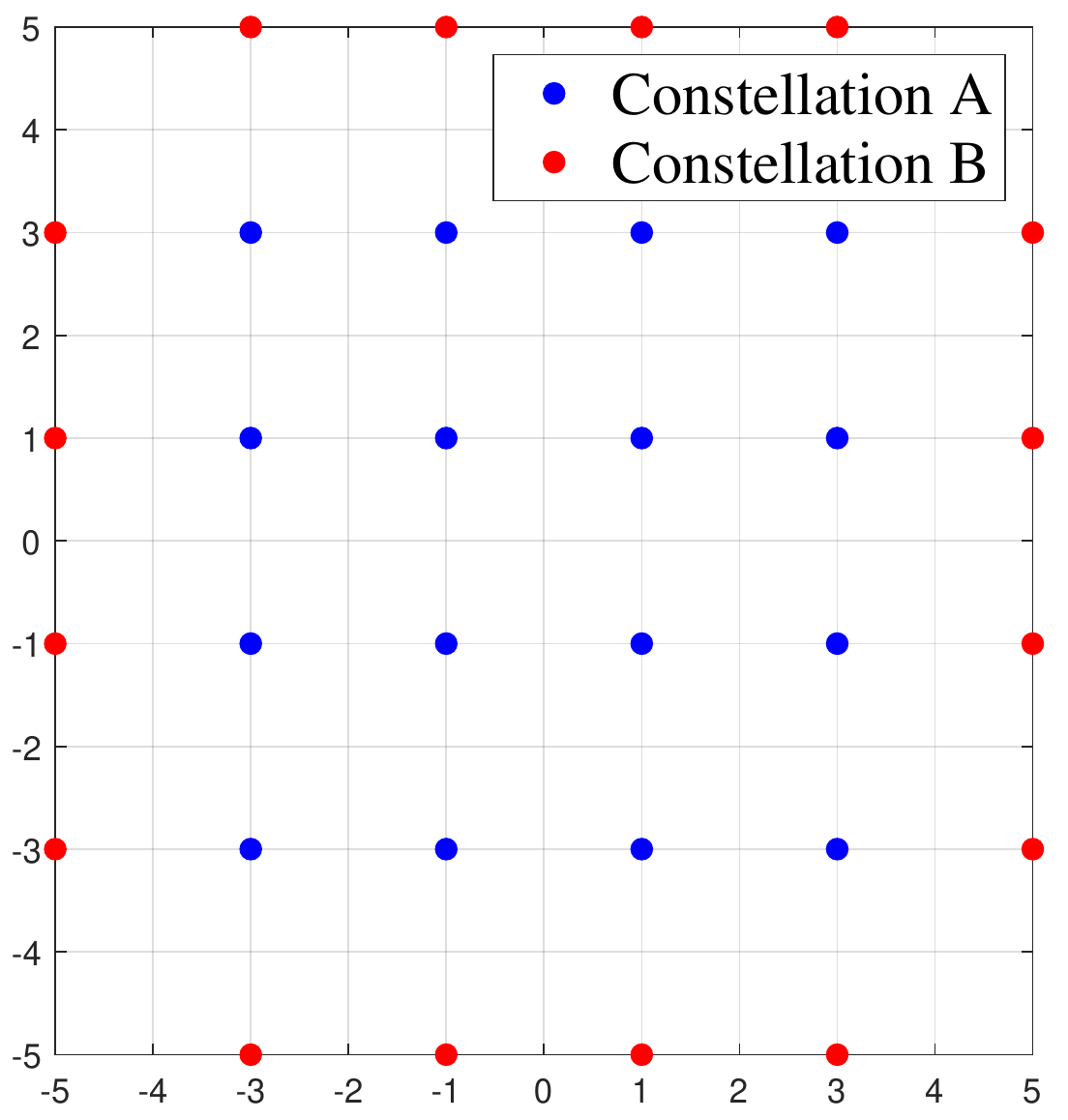}
  \caption{The conventional constellation pair $\mathcal{M}_A^{\mathrm{Conv,16QAM}}$ (Constellation A) and $\mathcal{M}_B^{\mathrm{Conv,16QAM}}$ (Constellation B) when $M_A = M_B = 16$.}\label{fig:CQAM}
\end{figure}

In \cite{mao2017dual} and \cite{mao2017generalized}, the two constellation pairs for $M_A=M_B=4$ and $M_A=M_B=16$ are proposed based on the classical constellations, quadrature phase shift keying (QPSK) and 16 quadrature amplitude modulation (16QAM), respectively. Specifically, $\mathcal{M}_A^{\mathrm{Conv,QPSK}} = \{\pm1\pm j\}$ and $\mathcal{M}_B^{\mathrm{Conv,QPSK}} = \{-(1+\sqrt{3}),-(1+\sqrt{3})j,(1+\sqrt{3}),(1+\sqrt{3})j\}$.
Also, the conventional constellation pair for $M_A=M_B=16$ is $\mathcal{M}_A^{\mathrm{Conv,16QAM}} = \{\pm1\pm j,\pm1\pm3j,\pm3\pm j,\pm3\pm3j\}$ and $\mathcal{M}_B^{\mathrm{Conv,16QAM}} = \{-3\pm5j,-1\pm5j,+1\pm5j,+3\pm5j,\pm5-3j,\pm5-1j,\pm5+1j,\pm5+3j\}$.

Figs. \ref{fig:CQPSK} and \ref{fig:CQAM} show the conventional constellation pair $(\mathcal{M}_A^{\mathrm{Conv,QPSK}},\mathcal{M}_B^{\mathrm{Conv,QPSK}})$ and $(\mathcal{M}_A^{\mathrm{Conv,16QAM}},\mathcal{M}_B^{\mathrm{Conv,16QAM}})$, respectively.

\section{The Proposed Signal Constellation Pair for DM-OFDM-IM}
WLOG, we omit the group index $g$ for convenience hereafter.
The well-known conditional pairwise error probability (CPEP) expression for the model in (\ref{eq:YXHW}) is given as \cite{jafarkhani2005space}
\begin{equation}\label{eq:PEP}
\Pr\left(\mathbf{X}\rightarrow \hat{\mathbf{X}}|\mathbf{H}\right) = Q\left(\frac{\delta}{N_0}\right) = Q\left(\frac{\delta\cdot SNR}{E_b}\right)
\end{equation}
where $\delta=\norm{\mathbf{X}\mathbf{H}- \hat{\mathbf{X}}\mathbf{H}}_2$, $SNR = E_b/N_0$, and $E_b$ is the energy per bit value. It is clear that in (\ref{eq:PEP}), for a fixed $SNR$, $\delta/E_b$ is the metric determining the CPEP.

In DM-OFDM-IM (or OFDM-IM), the error event can be classified into two types. One is the symbol demodulation error, where the modulated symbols are erroneously estimated even though $I_A^g$ is correctly estimated.
The other type is the index demodulation error, where $I_A^g$ is erroneously estimated. Clearly, the index demodulation error affects not only the bit stream $\mathbf{b}_1$ but also $\mathbf{b}_2$.
Since the worst case PEP event dominates the system performance, let us consider the worst case scenario for each error type.

First, the value of $\delta$ in the worst case of the symbol demodulation error, denoted as $\delta_1$, becomes
\begin{align*}
  \delta_1 = \min\bigg(&\min_{s_{A}\neq \hat{s}_{A},s_{A},\hat{s}_{A}\in \mathcal{M}_A} |H(\alpha_0)||s_A - \hat{s}_A|,\\
   &\min_{s_{B}\neq \hat{s}_{B},s_{B},\hat{s}_{B}\in \mathcal{M}_B}|H(\alpha_1)||s_{B} -\hat{s}_{B}|\bigg)
\end{align*}
where $\alpha_0 \neq \alpha_1$ and $\alpha_0, \alpha_1 = 1,\cdots,n$.
Since $|H(\alpha_0)|$ and $|H(\alpha_1)|$ have the same distributions, the desirable design criterion is
\begin{equation*}
  \min_{s_{A}\neq \hat{s}_{A},s_{A},\hat{s}_{A}\in \mathcal{M}_A} |s_A - \hat{s}_A| = \min_{s_{B}\neq \hat{s}_{B},s_{B},\hat{s}_{B}\in \mathcal{M}_B}|s_{B} -\hat{s}_{B}|
\end{equation*}
which means the typical criterion that the minimum Euclidean distances of two distinct signal points in constellations $\mathcal{M}_A$ and $\mathcal{M}_B$ have to be the same.

Second, the value of $\delta$ in the worst case of the index demodulation error, denoted as $\delta_2$, becomes
\begin{equation*}
  \delta_2 = \min_{s_{A}\in \mathcal{M}_A, s_B\in \mathcal{M}_B} \sqrt{ |H(\alpha_0)|^2 + |H(\alpha_1)|^2}|s_{A} - s_{B}|.
\end{equation*}
Therefore, $\min_{s_{A}\in \mathcal{M}_A, s_B\in \mathcal{M}_B}|s_{A} - s_{B}|$ has to be large. It can be seen that the diversity order of this error event is two.

Now, we will investigate the values of $\delta_1$ and $\delta_2$ of the conventional constellation pair and the proposed constellation pair in the following subsections.
\subsection{The Conventional Constellation Pair}
In Figs. \ref{fig:CQPSK} and \ref{fig:CQAM}, it is easily seen that
\begin{align*}
&\delta_1^\mathrm{Conv}=2|H(\alpha_0)|\\
&\delta_2^\mathrm{Conv} = 2\sqrt{ |H(\alpha_0)|^2 + |H(\alpha_1)|^2}.
\end{align*}
Note that the energy per bit values in Figs. \ref{fig:CQPSK} and \ref{fig:CQAM} are $E_b^\mathrm{Conv,QPSK} \simeq 1.8928$ and $E_b^\mathrm{Conv,16QAM} \simeq 4.444$, respectively.

\subsection{The Proposed Constellation Pair}
\begin{figure}
  \centering
  \includegraphics[width=.7\linewidth]{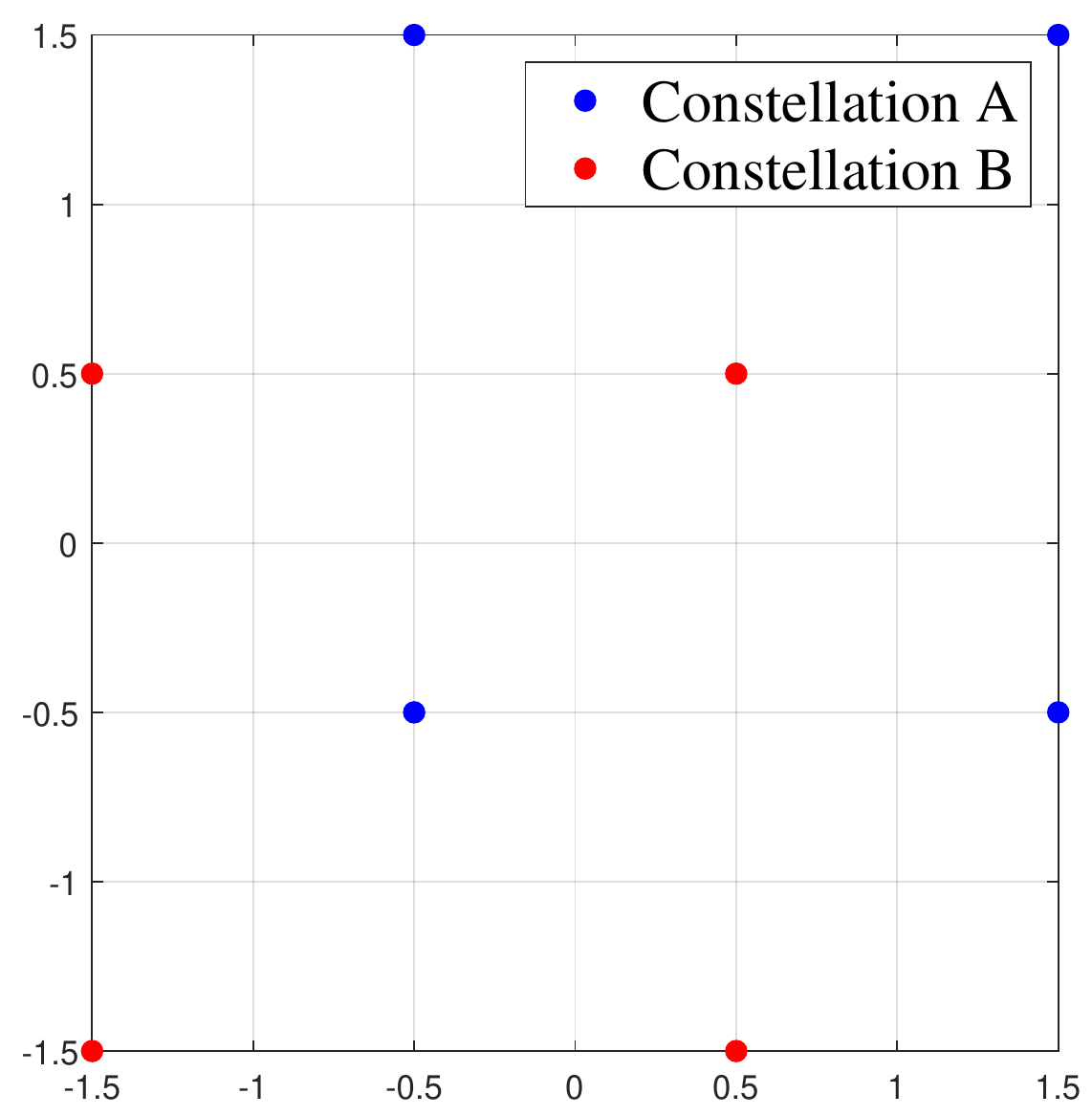}
  \caption{The proposed constellation pair $\mathcal{M}_A^{\mathrm{Prop,QPSK}}$ (Constellation A) and $\mathcal{M}_B^{\mathrm{Prop,QPSK}}$ (Constellation A) when $M_A = M_B = 4$.}\label{fig:PQPSK}
\end{figure}
\begin{figure}
  \centering
  \includegraphics[width=.7\linewidth]{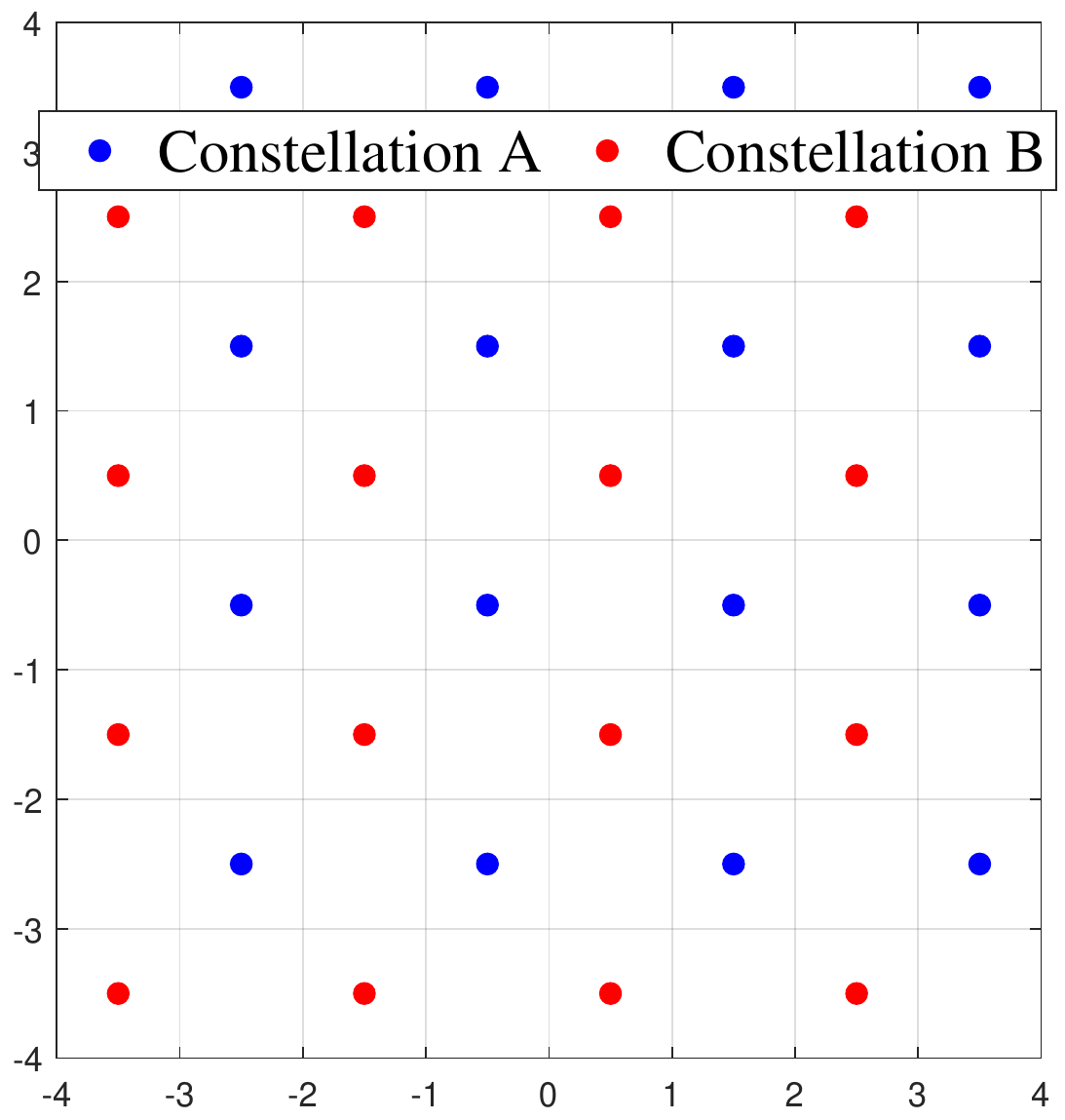}
  \caption{The proposed constellation pair $\mathcal{M}_A^{\mathrm{Prop,16QAM}}$ (Constellation A) and $\mathcal{M}_B^{\mathrm{Prop,16QAM}}$ (Constellation A) when $M_A = M_B = 16$.}\label{fig:PQAM}
\end{figure}

Based on the above PEP analysis, we propose the constellations for $M_A=M_B=4$ as
\begin{align*}
  &\mathcal{M}_A^{\mathrm{Prop,QPSK}} = \mathcal{M}_A^{\mathrm{Conv,QPSK}} \oplus 0.5 + 0.5j\\
   &\mathcal{M}_B^{\mathrm{Prop,QPSK}} = \mathcal{M}_A^{\mathrm{Conv,QPSK}} \oplus - 0.5 - 0.5j
\end{align*}
where $\oplus$ is the element-wise addition.
In the case of $M_A=M_B=16$, we propose the constellations as
\begin{align*}
  &\mathcal{M}_A^{\mathrm{Prop,16QAM}} = \mathcal{M}_A^{\mathrm{Conv,16QAM}} \oplus 0.5 + 0.5j\\
  &\mathcal{M}_B^{\mathrm{Prop,16QAM}} = \mathcal{M}_A^{\mathrm{Conv,16QAM}} \oplus - 0.5 - 0.5j.
\end{align*}
Figs. \ref{fig:PQPSK} and \ref{fig:PQAM} show the proposed constellation pair $(\mathcal{M}_A^{\mathrm{Prop,QPSK}},\mathcal{M}_B^{\mathrm{Prop,QPSK}})$ and $(\mathcal{M}_A^{\mathrm{Prop,16QAM}},\mathcal{M}_B^{\mathrm{Prop,16QAM}})$, respectively.

In Figs. \ref{fig:PQPSK} and \ref{fig:PQAM}, it is seen that
\begin{align*}
&\delta_1^\mathrm{Prop}=2|H(\alpha_0)|\\
&\delta_2^\mathrm{Prop} = \sqrt{2}\sqrt{ |H(\alpha_0)|^2 + |H(\alpha_1)|^2}.
\end{align*}
Note that the energy per bit values in Figs. \ref{fig:PQPSK} and \ref{fig:PQAM} are $E_b^\mathrm{Prop,QPSK} = 1$ and $E_b^\mathrm{Prop,16QAM} \simeq 2.3333$, respectively.

Plugging $\delta_1^\mathrm{Conv}$, $\delta_2^\mathrm{Conv}$, $\delta_1^\mathrm{Prop}$ ,$\delta_2^\mathrm{Prop}$, $E_b^\mathrm{Conv,QPSK}$, $E_b^\mathrm{Conv,16QAM}$, $E_b^\mathrm{Prop,QPSK}$, and $E_b^\mathrm{Prop,16QAM}$ into (\ref{eq:PEP}), we have the conclusion that the proposed constellation pair gives lower CPEP values than those from the conventional constellation pair for both QPSK and 16QAM cases.

It is remarkable that the proposed constellation pair has the same values of $\delta_1$ and $\delta_2$ as the values of the OFDM-IM while the $E_b$ values are slightly higher than those of OFDM-IM, where classical QPSK or 16QAM is used for active subcarriers in the OFDM-IM system.

\section{The Proposed Bit Mapping for DM-OFDM-IM}\label{sec:PDM-OFDM-IM}
In this section, we propose the new bit mapping scheme. The catastrophic effect of the index demodulation error incurring burst of errors can be relieved by the new bit mapping scheme. To the authors' best knowledge, the proposed approach here is the first work to solve that problem, which is the endemic drawback of the index modulation systems.

First, the proposed bit mapping scheme needs the constraint as
\begin{equation}\label{eq:const}
  M_A = M_B =M
\end{equation}
which is reasonable in a practical sense.
We consider the bit stream for generating $g$-th group as
\begin{align*}
  \mathbf{b} =& [b(1)~b(2)~\cdots~b(p)]\\
             =& [\mathbf{b}_1~\mathbf{b}_{2}]
\end{align*}
where
\begin{align*}
  &\mathbf{b}_1 = [b(1)~\cdots~b(p_1)] \\
  &\mathbf{b}_2 = [b(p_1+1)~\cdots~b(p_1+p_2)]
\end{align*}
whose sizes are $p_1$ and $p_2$, respectively. As in Section \ref{sec:DM-OFDM-IM}, the bit stream $\mathbf{b}_1$ is used to determine the index set $I_A$.

However, different from the conventional DM-OFDM-IM, the bit stream $\mathbf{b}_2$ are represented by $n$ bit substream with size $\log_2 M$ as
\begin{equation*}
  \mathbf{b}_2 = [\mathbf{b}_{2,1} \cdots \mathbf{b}_{2,n}]
\end{equation*}
where
\begin{equation*}
  \mathbf{b}_{2,\alpha} = [b(p_1 + (\alpha-1)\log_2 M + 1)~\cdots~b(p_1 + \alpha \log_2 M)]
\end{equation*}
and $\alpha = 1,\cdots,n$. Then the bit substream $\mathbf{b}_{2,\alpha}$ with size $\log_2 M$ can be mapped into a symbol in both $\mathcal{M}_A$ and $\mathcal{M}_B$ because of (\ref{eq:const}).

Now, considering the subcarrier pattern $I_A$, the $g$-th group $\mathbf{X}$ can be formed by, $\alpha = 1,\cdots,n$,
\begin{align}\label{eq:Pbs}
  X(\alpha) =\begin{cases}
  \mathcal{M}_A(\mathbf{b}_{2,\alpha}) & \alpha \in I_A\\
  \mathcal{M}_B(\mathbf{b}_{2,\alpha}) & \alpha \in I_B
  \end{cases}
\end{align}
where $\mathcal{M}_A(\cdot)$ and $\mathcal{M}_B(\cdot)$ mean the bits-to-symbol mapper using $\mathcal{M}_A$ and $\mathcal{M}_B$, respectively. The same procedure as in Section \ref{sec:DM-OFDM-IM} can be done up to (\ref{eq:ML}).

At the receiver, the estimate $\hat{\mathbf{X}}$ using ML detector in (\ref{eq:ML}) is obtained and easily $\hat{\mathbf{b}}_1$ is obtained from $\hat{I}_A$. Then the estimated bit stream $\hat{\mathbf{b}}_2$ can be obtained by
\begin{align}\label{eq:Psb}
  \hat{\mathbf{b}}_{2,\alpha} =\begin{cases}
  \mathcal{M}_A^{-1}(\hat{X}(\alpha)) & \alpha \in \hat{I}_A\\
  \mathcal{M}_B^{-1}(\hat{X}(\alpha)) & \alpha \in \hat{I}_B.
  \end{cases}
\end{align}

\subsection{Toy Example}

In this subsection, a simple example is given to understand the benefit of the proposed bit mapping scheme.
Let us consider the case $n=4, k=2$, and $M = 4$. Also, we assume the following situations;
\begin{align*}
  &I_A = \{1,3\}\\
  &\mathbf{b}_1 = [1 0]\\
  &\mathbf{b}_2 = [1 0 1 1~ 0 0 0 0~ 1 1 1 1~ 0 1 1 1].
\end{align*}
If the conventional bit mapping in Section \ref{sec:DM-OFDM-IM} is used, the result signal becomes
\begin{align*}
  &S_A = \{\mathcal{M}_A(1 0 1 1),\mathcal{M}_A(0 0 0 0)\}\\
  &S_B = \{\mathcal{M}_B(1 1 1 1),\mathcal{M}_B(0 1 1 1)\}\\
  &\mathbf{X}^{\mathrm{Conv}} = [\mathcal{M}_A(1 0 1 1)~\mathcal{M}_B(1 1 1 1)~\mathcal{M}_A(0 0 0 0)~\mathcal{M}_B(0 1 1 1)].
\end{align*}
However, the proposed bit mapping in (\ref{eq:Pbs}) gives us
\begin{equation*}
    \mathbf{X}^{\mathrm{Prop}} = [\mathcal{M}_A(1 0 1 1)~\mathcal{M}_B(0000)~\mathcal{M}_A(1111)~\mathcal{M}_B(0 1 1 1)]
\end{equation*}

If the subcarrier pattern $I_A$ is correctly detected at the receiver, there is no difference between the conventional and the proposed bit mapping schemes.
However, we assume that the subcarrier pattern is erroneously detected at the receiver as
\begin{align*}
  &\hat{I}_A = \{1,2\}.
\end{align*}
For convenience, we also assume that there is no AWGN.

Then, with $\mathbf{X}^{\mathrm{Conv}}$, the conventional DM-OFDM-IM receiver in Section \ref{sec:DM-OFDM-IM} gives
\begin{align*}
    &\hat{S}_A = \{\mathcal{M}_A(1 0 1 1),\mathcal{M}_B(1 1 1 1)\}\\
  &\hat{S}_B = \{\mathcal{M}_A(0 0 0 0),\mathcal{M}_B(0 1 1 1)\}\\
  &\hat{\mathbf{b}}_2^{\mathrm{Conv}} = [1011~\mathcal{M}_A^{-1}(\mathcal{M}_B(1111))~\mathcal{M}_B^{-1}(\mathcal{M}_A(0000))~0111].
\end{align*}
However, using the proposed bit demapping in (\ref{eq:Psb}), we have
\begin{align*}
  \hat{\mathbf{b}}_2^{\mathrm{Prop}} = [1011~\mathcal{M}_A^{-1}(\mathcal{M}_B(0000))~\mathcal{M}_B^{-1}(\mathcal{M}_A(1111))~0 1 1 1]
\end{align*}
while

If we use the proposed constellation pair in Figs. \ref{fig:PQPSK} or \ref{fig:PQAM} and the same gray mapping structure for $\mathcal{M}_A$ and $\mathcal{M}_B$, the bit difference between the input bit stream and the output bit stream of $\mathcal{M}_B^{-1}(\mathcal{M}_A(\cdot))$ or $\mathcal{M}_A^{-1}(\mathcal{M}_B(\cdot))$ is small.
Therefore, bit difference between $\mathbf{b}_2$ and $\hat{\mathbf{b}}_2^{\mathrm{Prop}}$ is lower than the bit difference between $\mathbf{b}_2$ and $\hat{\mathbf{b}}_2^{\mathrm{Conv}}$.

\section{Simulation Results}
To evaluate the benefit of the two proposed features, we simulate DM-OFDM-IM systems over a Rayleigh frequency selective fading channel, where we use $H^g(\alpha) \sim \mathcal{CN}(0,1)$ for all $g$ and $\alpha$. In simulations of DM-OFDM-IM systems, we use $N=128$, $n=4$, and $k=2$ is used. The subcarrier pattern $I_{A} \in \{\{1,2\},\{3,4\},\{1,3\},\{2,4\}\}$ is used, which corresponds to the bit stream $\mathbf{b}_1 \in \{00,11,10,01\}$ and thus $p_1=2$. The low complexity ML detection is used in all simulations.
\begin{figure}
  \centering
  \includegraphics[width=.99\linewidth]{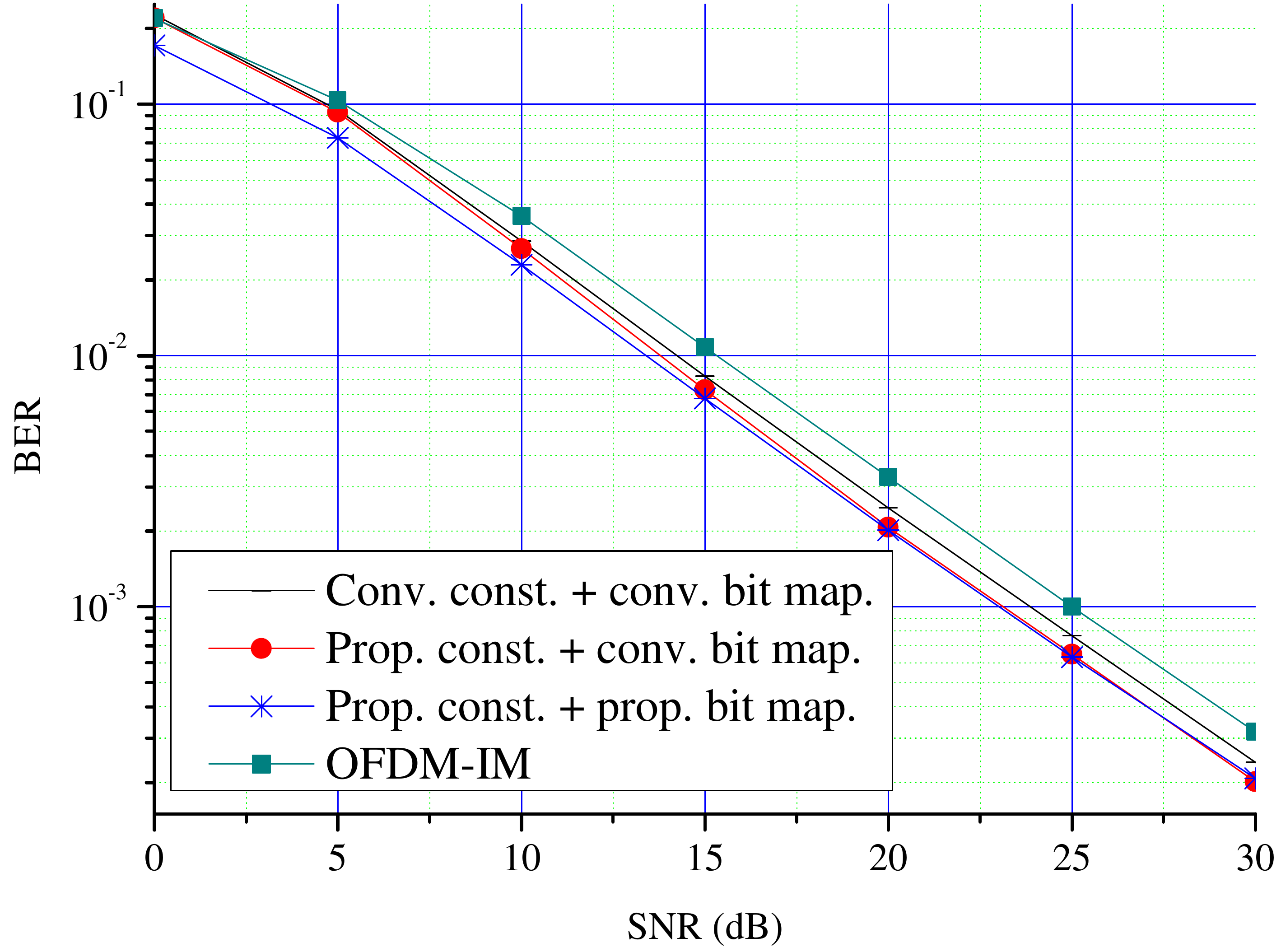}
  \caption{The BER performance comparison of DM-OFDM-IM when $M_A=M_B=4$ and OFDM-IM. The conventional constellation means DM-OFDM-IM with $(\mathcal{M}_A^{\mathrm{Conv,QPSK}}, \mathcal{M}_B^{\mathrm{Conv,QPSK}})$. The proposed constellation means DM-OFDM-IM with $(\mathcal{M}_A^{\mathrm{Prop,QPSK}},\mathcal{M}_B^{\mathrm{Prop,QPSK}})$.}\label{fig:QPSK}
\end{figure}

Fig. \ref{fig:QPSK} shows the BER performances of the DM-OFDM-IM systems with $M_A=M_B=4$ and OFDM-IM as benchmark. The legend means the follows;
\begin{itemize}
  \item \textbf{Conv. const. + conv. bit map.}: DM-OFDM-IM with $(\mathcal{M}_A^{\mathrm{Conv,QPSK}},\mathcal{M}_B^{\mathrm{Conv,QPSK}})$ and the conventional bit mapping.
  \item \textbf{Prop. const. + conv. bit map.}: DM-OFDM-IM with $(\mathcal{M}_A^{\mathrm{Prop,QPSK}},\mathcal{M}_B^{\mathrm{Prop,QPSK}})$ and the conventional bit mapping.
  \item \textbf{Prop. const. + prop. bit map.}: DM-OFDM-IM with $(\mathcal{M}_A^{\mathrm{Prop,QPSK}},\mathcal{M}_B^{\mathrm{Prop,QPSK}})$ and the proposed bit mapping.
  \item \textbf{OFDM-IM}: OFDM-IM in \cite{bacsar2013orthogonal}, where $N=128$, $n=4$, and $k=2$ are used and two active subcarriers are modulated by 16QAM.
\end{itemize}
Note that the OFDM-IM has the same spectral efficiency as the DM-OFDM-IM cases with 2.5 bits/s/Hz. It is shown that the proposed constellation pair gives a better BER performance than the conventional constellation pair.
Also, DM-OFDM-IM with the proposed constellation pair and the proposed bit mapping gives the best BER performance.

\begin{figure}
  \centering
  \includegraphics[width=.99\linewidth]{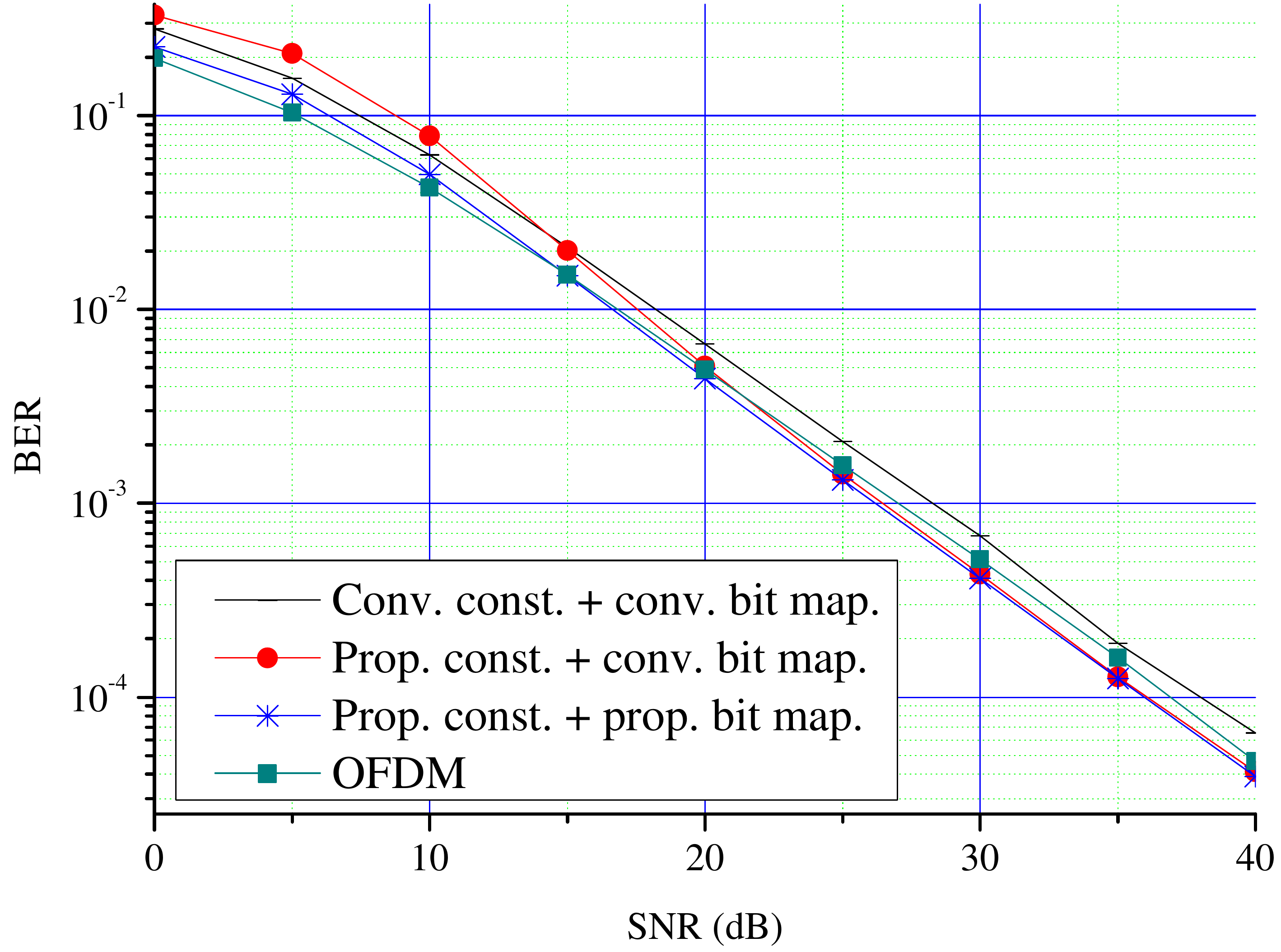}
  \caption{The BER performance comparison of DM-OFDM-IM when $M_A=M_B=16$ and OFDM. The conventional constellation means DM-OFDM-IM with $(\mathcal{M}_A^{\mathrm{Conv,16QAM}},\mathcal{M}_B^{\mathrm{Conv,16QAM}})$. The proposed constellation means the DM-OFDM-IM with $(\mathcal{M}_A^{\mathrm{Prop,16QAM}},\mathcal{M}_B^{\mathrm{Prop,16QAM}})$.}\label{fig:QAM}
\end{figure}

Fig. \ref{fig:QAM} shows the BER performances of the DM-OFDM-IM systems with $M_A=M_B=16$ and classical OFDM as benchmark. The legend means the follows;
\begin{itemize}
  \item \textbf{Conv. const. + conv. bit map.}: DM-OFDM-IM with $(\mathcal{M}_A^{\mathrm{Conv,16QAM}},\mathcal{M}_B^{\mathrm{Conv,16QAM}})$ and the conventional bit mapping.
  \item \textbf{Prop. const. + conv. bit map.}: DM-OFDM-IM with $(\mathcal{M}_A^{\mathrm{Prop,16QAM}},\mathcal{M}_B^{\mathrm{Prop,16QAM}})$ and the conventional bit mapping.
  \item \textbf{Prop. const. + prop. bit map.}: DM-OFDM-IM with $(\mathcal{M}_A^{\mathrm{Prop,16QAM}},\mathcal{M}_B^{\mathrm{Prop,16QAM}})$ and the proposed bit mapping.
  \item \textbf{OFDM}: OFDM, where the number of subcarriers is four and subcarriers are modulated by 16QAM.
\end{itemize}
Note that the DM-OFDM-IM systems have the spectral efficiency 4.5 bits/s/Hz and the classical OFDM has the spectral efficiency 4 bits/s/Hz.

In Fig. \ref{fig:QAM}, it is shown that the proposed constellation pair gives a better BER performance than the conventional constellation pair except in low SNR regime when the conventional bit mapping is used for both cases.
This is because the signal points in Fig. \ref{fig:PQAM} are regionally mixed. For example, the signal point $1.5+1.5j \in \mathcal{M}_A^{\mathrm{Prop,16QAM}}$ is surrounded by the four signal points $\{0.5+0.5j,0.5+2.5j,2.5+0.5j,2.5+2.5j\}\in \mathcal{M}_B^{\mathrm{Prop,16QAM}}$. This means that the signal point $1.5+1.5j$ could be easily confused with those four signal points because any direction of a noise vector can cause index error events. However, the signal points in Fig. \ref{fig:CQAM} are regionally separated. For example, the signal point $3+3j \in \mathcal{M}_A^{\mathrm{Conv,16QAM}}$ is surrounded by only two signal points $\{5+3j,3+5j\} \in \mathcal{M}_B^{\mathrm{Conv,16QAM}}$. In this case, some direction of a noise vector such as $-1-j$ does not cause error events. This characteristic is not revealed when we deal with the metric $\delta_1,\delta_2$ and this causes the worse BER performance of the proposed constellation pair than the conventional constellation pair in low SNR regime.

DM-OFDM-IM with the proposed constellation pair and the proposed bit mapping gives the best BER performance because the catastrophic effect of the index demodulation error is relieved by the new bit mapping scheme. Also, the DM-OFDM-IM shows a better BER performance compared to the OFDM system in high SNR region even though the proposed DM-OFDM-IM has the higher spectral efficiency.

\section{Conclusions}
In this paper, the well designed constellation pair for DM-OFDM-IM is proposed based on the PEP analysis. The DM-OFDM-IM system with the proposed constellation pair gives the better BER performance than that with the conventional constellation pair for almost SNR regime.
In addition, the new bit mapping scheme for DM-OFDM-IM is proposed, where the symbols are modulated from the bit stream in order from the front. As a result, the proposed bit mapping scheme can relieve the catastrophic effect of the index demodulation error incurring burst of errors, which is the endemic problem of the index demodulation systems. The simulation results show that the proposed constellation pair and the proposed bit mapping scheme substantially enhance the BER performance of the DM-OFDM-IM than the conventional DM-OFDM-IM.

\bibliographystyle{IEEEtran}
\bibliography{biblio,IEEEfull}

\end{document}